\documentclass[preprint,aps,showpacs]{revtex4}
\usepackage [cp1251]{inputenc}
\usepackage{amsmath}
\usepackage{bm}
\usepackage{graphicx}

\usepackage{color}
\usepackage{epsfig}

\newcommand{\nix}[1]{}

\newcommand{\be}{\begin{equation}}
\newcommand{\ee}{\end{equation}}
\newcommand{\bea}{\begin{eqnarray}}
\newcommand{\eea}{\end{eqnarray}}

\graphicspath{{pics/}}

\begin{document}

\title{Spin-dependent recombination in GaAs$_{1-x}$N$_x$ alloys at oblique
magnetic field}
\author{E.L. Ivchenko\footnote{ivchenko@coherent.ioffe.ru}, L.A. Bakaleinikov, M.M. Afanasiev, and V.K. Kalevich}
\affiliation{Ioffe Physical-Technical Institute, 194021 St. Petersburg, Russia}
\begin{abstract}
We have studied experimentally and theoretically the optical orientation and spin-dependent Shockley--Read--Hall 
recombination in a semiconductor in a magnetic field at an arbitrary angle $\alpha$ between the field and
circularly polarized exciting beam. The experiments are performed at room temperature in GaAs$_{1-x}$N$_x$ alloys where deep paramagnetic 
centers are responsible for the spin-dependent recombination. The observed magnetic-field dependences of the circular polarization $\rho({\bm B})$ and intensity $J({\bm B})$ of photoluminescence can be approximately described as a superposition of two Lorentzian contours, normal and inverted, with their half-widths differing by an order of magnitude. The normal (narrow) Lorentzian contour is associated with depolarization of the transverse (to the field) component of spin polarization of the localized electrons, whereas the inverted (broad) Lorentzian is due to suppression of the hyperfine interaction of the localized electron with the defect nucleus. The ratio between the height of one Lorentzian and depth of the other is governed by the field tilt angle $\alpha$. In contrast to the hyperfine interaction of a shallow-donor-bound electron with a large number of nuclei of the crystal lattice, in the optical orientation of the electron-nuclear system under study no additional narrow peak appears in the oblique field. This result demonstrates that in the GaAsN alloys the hyperfine interaction of the localized electron with the single nucleus of the paramagnetic center remains strong even at room temperature. For a theoretical description of the experiment, we have extended the theory of spin-dependent recombination via deep paramagnetic centers with the nuclear angular momentum $I = 1/2$ developed previously for the particular case of the longitudinal field. The calculated curves $\rho({\bm B})$, $J({\bm B})$ agree with the approximate description of the experimental dependences as a sum of two Lorentzians, and an additional narrow shifted peak does not appear in the computation as well. 
\end{abstract}
  \maketitle
\section{Introduction}
In recent years the spin-dependent Shockley--Read--Hall recombination 
attracts a considerable attention since it allows one to obtain
anomalously high values of spin polarization of conduction electrons
in nonmagnetic semiconductors at room temperature \cite{1, 2,Nature2009,3, 4, 5, 6} 
(see also the review paper \cite{JPCM2010}). The origin of this effect is the spin-dependent
capture of optically oriented conduction electrons onto deep paramagnetic centers in which case the electrons
localized on the centers become dynamically spin-polarized and,
acting as a spin filter, block the further recombination of conduction-band electrons with the
majority spin orientation. The spin filter efficiency increases with increasing the pumping
which allows one, at high pumping powers, to get the electron polarization close to 100\%.

The nonlinear coupling of the spin subsystems of free and localized electrons
leads to a number of striking effects in a magnetic field. Particularly, the electron spin depolarization in the magnetic field 
perpendicular to the exciting beam (Hanle effect, the Voigt geometry) is described by a superposition of two Lorentzian
contours with widths at half maximum differing by two or three orders: the large spin relaxation time of
localized electrons ($\sim$1 ns) determines the width of the narrow
contour ($\sim$100\,G) whereas the short lifetime of free electrons
($\sim$1\,ps) sets the width $\sim$25 kG of the wide contour
\cite{JPCM2010,PhysicaB2009}.

Also, it has recently been established 
\cite{PRB2012,JETPLett2012,Sweden2013,Toulouse2014,PRB2015} that the magnetic
field directed along the exciting beam (the Faraday geometry)
can lead to an increase in the efficiency of spin filter and,
as a consequence, to a substantial (up to twice) enhancement
of the electron polarization and intensity of the edge photoluminescence
(PL) at low and moderate pumping rates. This effect is based on
the longitudinal-magnetic-field induced suppression of the electron spin depolarization
caused by the hyperfine interaction of a localized electron with the nucleus of the paramagnetic
center which localizes this electron. Additionally, the experiment shows that the magnetic-field dependences
of PL circular polarization and intensity are shifted with respect to zero
field by $\sim$100\,G \cite {PRB2012, JETPLett2012,Sweden2013}. This shift changing its sign with the sign reversal of the pump circular polarization 
has been attributed \cite{JETPLett2012,Sweden2013,Toulouse2014} to the Overhauser field $B_N$ created by the dynamically polarized nucleus of the paramagnetic center 
and acting back on the localized electron. In Refs. \cite{JETPLett2012, Sweden2013,Toulouse2014,PRB2015} 
the analysis of the experimental data obtained at room temperature
was performed assuming the regime of strong hyperfine coupling of the electron spin with the
nuclear spin of the paramagnetic center on which the electron is localized. 
By definition, in this regime the hyperfine splitting between the levels
with angular momenta $ I + 1/2 $ and $ I - 1/2 $ ($ I $ is the nuclear momentum) exceeds their widths defined by the 
electron and nuclear reciprocal spin relaxation lifetimes. The strong coupling regime was proved by the observation of a multiline spectrum 
of the optically detected electron spin resonance (EPR) on the Ga$^{2+}$ self-interstitial defects responsible for the
spin-dependent recombination in GaAsN \cite{Nature2009,3,Sweden2013}. However, the EPR measurements were performed at helium
temperature and the results of their analysis were extrapolated to room temperature. 

On the other hand, it is known \cite {OO, Dyak2008} that, in case of the hyperfine
coupling of a localized electron with a large number $\sim10^5$ of nuclei of the crystal lattice 
(weak coupling of the electron with each particular nucleus), the nuclear field can
reach a value of $B_N^{\rm max} \sim 5$ T. Being added to the external magnetic field, 
it leads to a radical change in the electron polarization. The action of nuclear field most clearly manifests itself in the external magnetic field tilted 
at an angle $\alpha$ with respect to the optical pumping direction. In this case the nuclear field $B_N(\alpha) \propto B_N^{\rm max} \cos{\alpha}$.
Therefore, a deviation of the external magnetic field from the sample surface plane by only a few degrees can result in the Overhauser field of several kilogauss which shifts away the Hanle curve by the same value from the zero-field point \cite{OO,Safarov1974,FTT1981}. Thus, the asymmetry of Hanle curve in a tilted magnetic field can be used for a 
qualitative determination of the type, strong or weak, of hyperfine coupling at the paramagnetic centers in GaAsN at room
temperature.

In this work we have performed room temperature measurements of 
the PL circular polarization and intensity in GaAsN
in the magnetic field oriented at different angles to the exciting
beam of circularly polarized light. Optical excitation is carried out by a laser
radiation normally incident on the sample surface; the secondary emission is recorded in the backscattering geometry.
It turns out that the narrow contour of the Hanle curve
associated with the depolarization of localized electrons
by the perpendicular component of the external field is not shifted from 
zero magnetic field even at high, up to 45$^{\circ}$, deviation angles of the magnetic field from the plane normal to the
excitation direction. This observation allows us to conclude that the hyperfine interaction at the paramagnetic centers in GaAsN
remains strong, i.e., at room temperature a localized electron also interacts mainly with the single nucleus of the paramagnetic center.
This result is a direct experimental proof of the strong
hyperfine coupling in dilute GaAsN alloys at room 
temperature, in the same conditions where the anomalously large electron spin polarization was registered.

For a theoretical description of the experimental results, we apply the model of optical orientation in semiconductors taking into account (i) the spin-dependent
recombination via deep paramagnetic centers and (ii) the hyperfine
coupling at the defects and developed earlier for the longitudinal
magnetic field and the spin $I$=1/2 \cite{PRB2015}. We have generalized this model to
arbitrary angle between the magnetic field and the exciting beam direction in which case the number of equations
increases from 11 to 25 as compared to the model in the longitudinal
field. The description becomes more complicated because, at oblique magnetic field, all the spin-density matrix components for the defects with one or two
electrons are different from zero. We have calculated the basic dependencies. The maxima of computed magnetic-field dependences of the PL 
circular polarization and intensity are not shifted away from the field zero, in the full agreement with the experimental curves.

The paper is organized as follows. Section 2 describes the experimental conditions and obtained results, Section 3 is devoted to a description of the theory for the nuclear spin $I$ =1/2 and the tilted magnetic
field, in Section 4 we discuss the results of calculation and compare them with the
experiment.
\section{Experimental conditions and results}
We investigated the electron spin polarization in the undoped
dilute GaAs$_{0.98}$N$_{0.02}$ alloy grown by molecular beam epitaxy (with a plasma
source for nitrogen) in the form of a 0.1\,$\mu$m-thick layer on the
semi-insulating GaAs(001) substrate \cite{JAP2005}. The spin polarization $P$ of the free electrons was generated under interband
absorption of the circularly polarized light \cite{OO}. 
A continuous-wave Ti:sapphire laser was used for PL excitation. The exciting laser beam was directed
normally to the sample surface (hereinafter the $z$ axis), the PL was registered in the backward direction by a photomultiplier with an InGaAsP
photocathode. We measured the stationary degree $\rho$ of circular
polarization of the edge PL proportional to the degree of free-electron polarization
\cite{OO}: $\rho = P 'P$, where the depolarization factor $P'\leq 1$. The degree of PL polarization is defined as
the ratio $\rho=(J^+ - J^-)/(J^+ + J^-)$, where $J^+$ and $J^-$ are the right ($\sigma^+$) and
left ($\sigma^-$) circularly polarized PL components, respectively, $J=J^+ + J^-$ is
the total PL intensity. The values of $J^+$ and $J^-$ were measured at
room temperature using a high-sensitive polarization analyzer \cite{Analyzer}.
\begin{figure}
\includegraphics[width=14.0cm,angle=0]{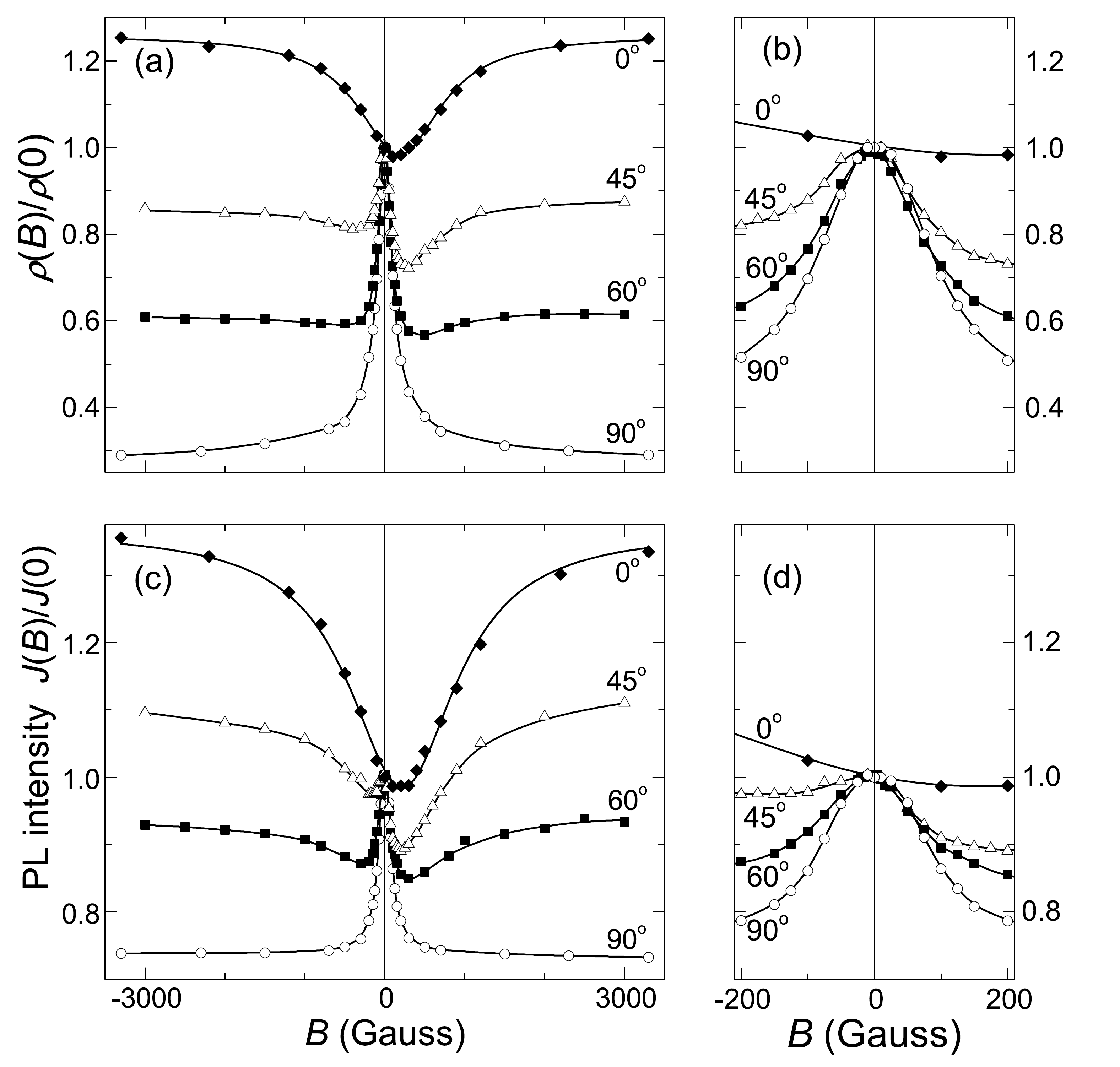}
\caption{\label{fig1} Magnetic-field dependences of the circular polarization degree (a, b) and intensity (c, d)
of photoluminescence of the GaAs$_{0.98}$N$_{0.02}$ alloy measured at the magnetic-field tilt angles $\alpha =
0,45^{\circ},60^{\circ}$and $90^{\circ}$. The pumping power $W$ = 60 mW, $T$ = 300 K. The solid curves are the guides for the eye.}
\end{figure}

Figure \ref{fig1} shows magnetic-field dependences of the edge PL intensity $J({\bm B})$ and degree of circular polarization $\rho({\bm B})$ 
measured in the GaAs$_{0.98}$N$_{0.02}$ alloy in the magnetic field tilted at the angle $\alpha = 0$, $45^{\circ}$, 
$60^{\circ}$ and $90^{\circ}$ with respect to the excitation beam. 

In the perpendicular field ($\alpha = 90^{\circ}$, circles) the values of $\rho({\bm B})$ and $J({\bm B})$
 rapidly decrease by tens percent with a half-width at half-maximum $B^{\perp}_{1/2} \sim 100$ G. The remaining intensity is independent
of the field, while the polarization slowly decreases with a further increase of the field. The strong changes in $\rho({\bm B})$ and $J({\bm B})$ within the interval of 
 100 G are consequences of the magnetic depolarization (Hanle effect) of electrons localized on deep paramagnetic centers \cite{1,PhysicaB2009,JPCM2010}.
In a strong transverse field  the localized electrons are unpolarized. Therefore the recombination rate of free photoelectrons
and hence the intensity of interband photoluminescence are independent of the field. A slow reduction of the PL circular polarization occurs due to manifestation 
of the Hanle effect on free electrons characterized by a small spin lifetime. For ${\bm B} \perp z$, in the whole range of fields the dependences $\rho({\bm B}), J({\bm B})$ are symmetric with respect to an inversion of the magnetic field.
\begin{figure}
\includegraphics[width=14.0cm,angle=0]{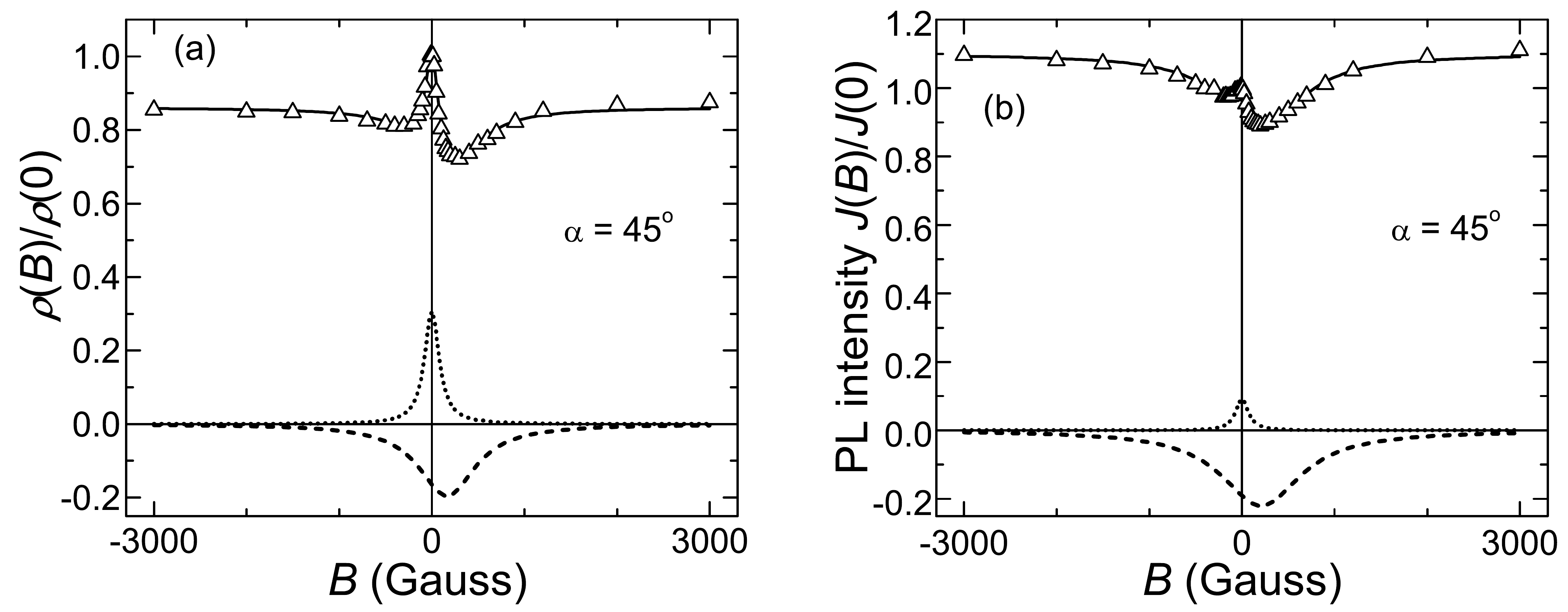}
\caption{\label{fig2} 
Description of the experimental dependences $\rho({\bm B})$ and $J({\bm B})$ (triangles) measured at the
tilt angle  $\alpha = 45^{\circ}$ and pump power 60 mW by a superposition of the normal (dotted) and inverted
(dashed line) Lorentzians.}
\end{figure}

In the longitudinal field ($\alpha=0$, diamonds) the PL polarization and intensity increase
as a result of suppression of spin relaxation of the localized electrons, and the increase is described with good accuracy by inverted 
Lorentzians with half-width at half minimum of $B^{\parallel}_{1/2} \sim 1$ kG \cite{PRB2012,JETPLett2012,Sweden2013,Toulouse2014,PRB2015}.
The curves $\rho(B)$ and $J(B)$ are asymmetric: their minima are shifted relative to the point $B=0$, and the shift changes its sign under reversal of the excitation 
circular polarization from $\sigma^+$ to $\sigma^-$. As shown in Refs.~\cite{PRB2012,JETPLett2012,Sweden2013,Toulouse2014} this shift is caused by the Overhauser field $B_N$.
It increases with the pump (up to $B^{\rm max}_N$$\approx$ 250 G \cite{JETPLett2012}) and, for the excitation power $W =$ 60 mW corresponding to Fig. \ref{fig1}, takes the value $\approx$~200 G.

In the oblique field for $\alpha = 45^{\circ}$ and $60^{\circ}$ (triangles and squares, respectively), the curves $\rho({\bm B})$ and $J({\bm B})$ can be approximately represented as a sum of a constant and a superposition of two Lorentzian contours, normal and inverted, with very different widths of $\sim$100 G and $\sim$1000 G, respectively. In Fig. \ref{fig2} these two contours are displayed by dotted and dashed lines. The narrow contour is symmetric with respect to the zero field while, in contrast, the broad one is shifted relative to the point $B=0$ in the same direction as it does in purely longitudinal geometry. One can see from Fig. \ref{fig1}, (a) and (c), that with increasing the tilt angle the contribution of the narrow Lorentzian increases and that of the broad Lorentzian decreases.  A big difference between the widths of the narrow and broad contours, as well as the change of their relative contributions with varying the angle $\alpha$, indicates that the shape of curves $\rho({\bm B})$ and $J({\bm B})$ in the oblique field is determined by the interplay between the depolarization of localized electrons by the transverse component of the external field
${\bm B}_{\perp}$ and the slowdown of their spin relaxation due to the longitudinal component $B_z$. It is noteworthy that the peak of narrow part of the curve $\rho({\bm B})$ or $J({\bm B})$ in Fig. \ref{fig1} does not shift with decreasing the tilt angle from 90$^{\circ}$ to 45$^{\circ}$ which is particularly evident in panels (b) and (d) of Fig. 1 where the field scale is especially stretched.

\section{Theory for the nucleus with $I = 1/2$}
The main purpose of the theoretical part of this paper is to demonstrate the absence of a 
sharp peak shifted from the point $B=0$ in the curves $\rho({\bm B})$ and $J({\bm B})$ in Fig. \ref{fig1}. In this section we
generalize the theory of spin-dependent recombination developed in \cite{PRB2015} for a nucleus with
the angular momentum $I$ = 1/2 and the longitudinal magnetic field ${\bm B} \parallel z$ and consider the
case of an arbitrary tilt angle $\alpha$ between ${\bm B}$ and the $z$ axis. The model with the nuclear spin 1/2 is
relatively simple and, as will be shown, qualitatively explains the symmetricity of the central
peak and confirms the regime of strong hyperfine interaction of the localized electron,
thereby excluding the regime of weak interaction. As stated in \cite{PRB2015}, in order to explain the shift of the curves $\rho({\bm B})$ and $J({\bm B})$ in the 
longitudinal field one should consider nuclei with the moment $I>1/2$, this task is planned to be addressed elsewhere.

Using the kinetic equations for the spin-density matrices for the deep centers with one electron and two electrons, we have derived the set of equations interrelating 25 variables $n, p, N_2, N_1, S_{\lambda}, S_{c,\lambda}, S_{n1, \lambda}, S_{n2, \lambda}, \Phi_{\lambda \eta}$ ($\lambda, \eta = x,y,z$) for an arbitrary direction of the field ${\bm B}$. These equations have the form
\begin{subequations} \label{Lall}
\begin{align}
&c_p N_2 p = G \:, \label{1a}\\
& c_n \left( n N_1 - 4{\bm S} {\bm S}_c \right) = G \:, \label{1b}\\
&c_n \left( N_1 {\bm S} - n {\bm S}_c \right) + \frac{\bm S}{\tau_s} + {\bm S} \times {\bm \omega} = \frac{P_i}{2} G {\bm o}_z\:, \label{1c}\\
& c_n \left( n S_{c \lambda} - N_1 S_{\lambda}\right) + \frac{S_{c \lambda}}{\tau_{sc}} + \left( {\bm S}_c \times {\bm \beta} \right)_{\lambda}
+ \Omega\ e_{\lambda \mu \nu} \Phi_{\mu \nu} = 0 \label{1d} \:, \\
& c_n \left( n \Phi_{\lambda \eta} - S_{\lambda} S_{n1,\eta} \right) + \left( \frac{1}{\tau_{sc}} + \frac{1}{\tau_{n1}} \right)
\Phi_{\lambda \eta} + e_{\lambda \mu \nu} \Phi_{\mu \eta} \beta_{\nu} + \frac{\Omega}{4} e_{\lambda \eta \mu} \left( S_{n1,\mu} - S_{c\mu} \right)= 0\:, \label{1e}\\
& c_n \left( n S_{n1,\lambda} - 4 S_{\mu} \Phi_{\mu \lambda} \right) + \frac{S_{n1,\lambda}}{\tau_{n1}}  - \Omega\ e_{\lambda \mu \nu}
\Phi_{\mu \nu}= c_p p S_{n2,\lambda} \:, \label{1f}\\
&c_p p S_{n2,\lambda} + \frac{S_{n2,\lambda}}{\tau_{n2}} = c_n \left( n S_{n1, \lambda} - 4 S_{\mu} \Phi_{\mu \lambda}\right) \:, \label{1g}\\
& N_1 + N_2 = N_c\:, \label{1h}\\
& p = n + N_2 \:. \label{1i}
\end{align}
\end{subequations}
Here $G$ is the optical generation rate of electrons (and holes) in the conduction band (valence band), ${\bm o}_z$ is the unit vector directed along the normal $z$, $e_{\lambda \eta \mu}$ is the antisymmetric unit tensor of the third rank of Levi-Civita, $P_i$ is the degree of spin orientation of photoelectrons at the time of excitation, $n$ and $p$ are the densities of free electrons in the conduction band and free holes in the valence band, ${\bm S}$ and ${\bm S}_c$ are the spin polarizations of free and localized electrons, $N_1, N_2$ and $S_{n1, \lambda}, S_{n2, \lambda}$ are the concentrations of deep centers and components of average total nuclear spin in the defect states with one and two bound electrons, $\Phi_{\lambda \eta}$ are components of the tensor describing the correlation between the electron spin and the spin of nucleus where this electron is localized. In the absence of correlation, $\Phi_{\lambda \eta} = S_{c \lambda} S_{n1, \eta}/N_1$; in case of the full spin orientation along the axis $z$, the component $\Phi_{zz} = N_1/4$ and all other components $\Phi_{\lambda \eta}$ with $\lambda \neq z,\eta \neq z$ vanish. Equations (\ref{Lall}) are derived from the equations for the spin-density matrices $\rho_{s',m';s,m}$ and $N_{2; m', m}$ for the defects with one or two electrons, where $s,s' = \pm 1/2$ and $m,m' = \pm 1/2$ are the electron and nuclear spin projections onto the $z$ axis. The sought physical quantities in Eqs.~(\ref{Lall}) are expressed in terms of the density matrix components as follows:
\begin{eqnarray}
&&N_2 = \sum\limits_{m} N_{2; m, m}\:,\:N_1 = \sum\limits_{sm} \rho_{s,m;s,m}\:, \nonumber\\
&&S_{c \lambda} = \frac12 \sum\limits_{ss'm} \sigma^e_{\lambda,ss'} \rho_{s',m;s,m}\:,\:
S_{n1, \lambda} = \frac12 \sum\limits_{smm'} \sigma^n_{\lambda,mm'} \rho_{s,m';s,m}\:,\nonumber\\
&&S_{n2, \lambda} = \frac12 \sum\limits_{mm'} \sigma^n_{\lambda,mm'} N_{2; m', m}\:, \nonumber\\
&&\Phi_{\lambda \eta} = \frac14 \sum\limits_{ss'mm'} \sigma^e_{\lambda,ss'} \rho_{s',m';s,m} \sigma^n_{\eta,mm'} \:,
\end{eqnarray}
where $\sigma^e_{\lambda,ss'}, \sigma^n_{\lambda,mm'}$ are the Pauli spin matrices for the electron and nucleus. Equations~(\ref{Lall}) contain the following set of system parameters: the density of deep centers (defects) $N_c$, the capture coefficient $c_n$ of a free electron onto a deep level with one localized electron, the coefficient $c_p$ of recombination of a free hole with one of the two electrons localized on one defect, the spin relaxation times of free electrons ($\tau_s$), bound electrons ($\tau_{sc}$), nuclei of defects with one bound electron ($\tau_{n1}$) and two bound electrons ($\tau_{n2}$), Land\'e factors for electrons in the conduction band, $g$, and for bound electrons, $g_c$, that determine the corresponding Larmor precession frequencies 
${\bm \omega} = g \mu_B {\bm B}/\hbar$, ${\bm \beta} = g_c \mu_B {\bm B}/\hbar$ ($\mu_B$ is the Bohr magneton). The frequency $\Omega$ is given by the ratio of $A/\hbar$, where $A$ is the constant of hyperfine interaction of the electron and nuclear spins described by Hamiltonian
\[
{\cal H}_{\rm hf} = A ~{\bm s}^e \cdot {\bm s}^n = A \left[ s^e_z s^n_z +
 \frac12 \left( s^e_+ s^n_- + s^e_- s^n_+ \right)\right]\:,
\]
$s^e_{\pm} = s^e_x \pm {\rm i} s^e_y$,~ $s^n_{\pm} = s^n_x \pm {\rm i} s^n_y$, ${\bm s}^e$ and ${\bm s}^n$ are spin operators with the components $\hat{\sigma}_{\lambda}^e/2$ and $\hat{\sigma}_{\lambda}^n/2$. As in the previous work \cite{PRB2015}, we neglect the direct action of the magnetic field on the nuclear spin. Note that, in the longitudinal magnetic field, the number of nonzero variables is reduced to 13, they are $n, p, N_2, N_1, S_z, S_{cz}$, $S_{n1, z}, S_{n2, z}, \Phi_{zz}$, $\Phi_{xy} = - \Phi_{yx}$ and $\Phi_{xx} = \Phi_{yy}$, among them 11 are linearly independent. In Ref.~\cite{PRB2015} we expressed $\Phi_{xy}, \Phi_{yx}$, $\Phi_{xx}$ and $\Phi_{yy}$ through the remaining 9 variables and assigned the number (29) to the obtained set of 9 equations. In the oblique field all the 25 values are different from zero, the set (\ref{Lall}) in general requires numerical solution and allows analytical solutions only in special limiting cases.

The numerical solution of the system (\ref{Lall}) is conveniently divided into three stages. First of all, we exclude the spin polarization ${\bm S}_c$ of localized electrons out of Eqs. (\ref{1b}) and (\ref{1c}). To do this, we multiply Eq. (\ref{1b}) by $n$, find the scalar product of Eq. (\ref{1c}) with $4{\bm S}$ and subtract one from the other. Solving the resulting equation for the concentration $N_1$, we find it as a function of the electron concentration $n$ and the polarization degree of free electrons, ${\bm P} = 2 {\bm S}/n$. Next we introduce the dimensionless quantities
\[
Y=\frac{N_2}{N_c}=\frac{N_c -N_1}{N_c}, \: \: Z=\frac{n}{N_c} \:, \: X=\frac{G}{c_p N_c ^2}
\]
and parameters ${\tau}_h^* = (c_p N_c)^{-1}$, $a=c_p /c_n$. Taking into account the relation (\ref{1i}) Eq.~(\ref{1a}) is reduced in the new variables to 
\begin{equation} \label{YZX}
Y (Y + Z) = X\:.
\end{equation}
Using the linear relationship (\ref{1h}) between $N_1$ and $N_2$ we express the dimensionless concentration of two-electron centers via the dimensionless concentration of free electrons
\begin{equation} \label{La2}
Y=\frac{L + M Z}{Z}\:, \:
L= - aX \frac{ 1-P_i P_z}{1-{\bm P}^2}\:, \: M=1 - a \frac{{\tau}_h^*}{\tau_s } \frac{{\bm P}^2}{1-{\bm P}^2}\:.
\end{equation}
Substituting the expression (\ref{La2}) for $Y$ into Eq. (\ref{YZX}) we find that, for a given pseudovector ${\bm P}$, the value of $Z$ satisfies the third-order equation and can be found analytically
by using Cardano's formula.

\begin{figure}
\includegraphics[width=14.0cm,angle=0]{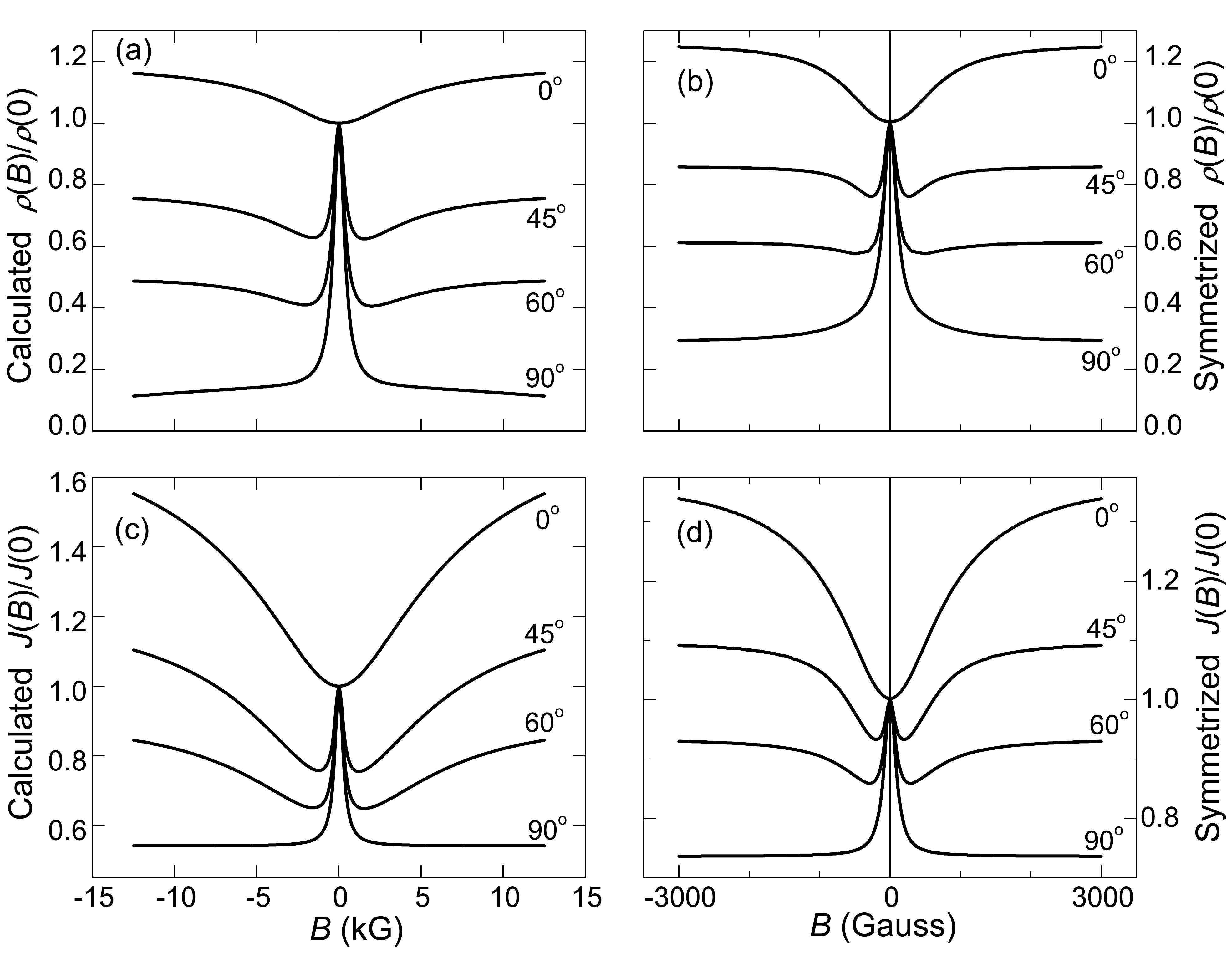}
\caption{\label{fig3} Magnetic-field dependences of the circular polarization degree (a) and intensity (c) of
photoluminescence calculated in the model with the nuclear angular momentum $I$ =1/2 for the
following parameters:  $\tau^* = (c_n N_c)^{-1}$= 2 ps, $\tau^*_h$ = 30 ps, $\tau_s$
= 140 ps, $\tau_{sc}$ = 700 ps, $\tau_{n1} = \tau_{n2}$ = 150\,ps,
$N_c$ = 3$\cdot$10$^{15}$ cm$^{-3}$, $g$ = 1, $g_c$ = 2, $A$ = 17
$\mu$eV, $P_i$ = 0.13, $W$ = 60 mW. The magnetic-field
tilt angles are indicated on the right side, near each curve. The similar curves in panels (b) and (d)
are obtained by symmetrization of the experimental dependences presented in Figs. \ref{fig1}(a) and \ref{fig1}(c).}
\end{figure}

The second stage is finding ${\bm S}_c$ from the equation set (\ref{1d})--(\ref{1g}). First, we exclude $S_{n2,\lambda}$ from Eq. (\ref{1f}) using Eq. (\ref{1g}).
Equations (\ref{1d})--(\ref{1f}) for given values of $n$, $N_1$ and ${\bm P}$ is a linear system with respect to the unknowns $S_{c \lambda}, S_{n1, \lambda}$ and $\Phi_{\lambda \eta}$ .
To solve it, one can use the standard procedure for the numerical solution of linear systems. Moreover, the matrix structure of the system allows the reduction of the problem to the successive solution of $3 \times 3$ systems of linear equations. Indeed, since the direct magnetic-field effect on the nuclear spin is neglected, each equation (\ref{1e}) contains the components $\Phi_{\lambda \eta} $ with the same index $\eta$. Therefore, for given values of $S_{c \lambda}, S_{n1, \lambda}$, one can solve nine equations (\ref{1e}) sequentially for sets $(\Phi_{xx} ,\Phi_{yx} ,\Phi_{zx}) $, $(\Phi_{xy} ,\Phi_{yy} ,\Phi_{zy})$ and $(\Phi_{xz} ,\Phi_{yz} ,\Phi_{zz})$. Since $\Phi_{\lambda \eta}$ are linearly dependent on $S_{c, \lambda}, S_{n1, \lambda} $, the substitution of found $\Phi_{\lambda \eta}$ into Eq. (\ref{1f}) gives a linear system for $S_{n1, \lambda}$. Substitution of its solution into (\ref{1d}) leads to a system for $ S_{c \lambda} $. Thus, the equation set (\ref{1d})--(\ref{1g}) allows one to find $ S_{c \lambda} $ as a function of ${\bm P}$. Substitution of this function into Eq.~(\ref{1c}) reduces the problem to a solution of a system of three nonlinear equations for the components $P_{\lambda}$. In the third stage, the solution of the latter system is found by using standard numerical procedure.
\section{Comparison with experiment and discussion}
In Fig. \ref{fig3}, (a) and (c), we present the results of calculation of the PL intensity $J \propto n p$ and the circular polarization degree 
\begin{equation} \label{p'}
\rho = \frac{2 P' S_z}{n} 
\end{equation}
carried out for four different angles $\alpha$ between the field $ {\bm B} $ and the $z$ axis. 
Here $P'$ is the depolarization factor \cite{JPCM2010}, the parameters used in the calculation are given in the caption to Fig. \ref{fig3}. The pumping power $W$ (in units of mW) is related to the optical generation rate $G$ entering Eqs. (\ref{Lall}) by $G = 7.5 \times 10^{23} W$ cm$^{-3}$s$^{-1}$. Calculations show that, for nuclei with $I = 1/2$, in an oblique magnetic field with $\alpha \neq 0$ and $\alpha \neq 90^{\circ}$ the dependences $\rho ({\bm B})$ and $J({\bm B})$ are almost insensitive to inversion of the field direction: the calculated degrees of asymmetry
\[
\rho_{\rm as}({\bm B}) = \frac{\rho({\bm B}) - \rho(-{\bm B})}{2}\:,\: J_{\rm as}({\bm B}) = \frac{J({\bm B}) - J(-{\bm B})}{2}
\]
do not exceed a few percent and are invisible in the scale of Fig. \ref{fig3}. Therefore, as mentioned above, in this work we have focused on the theoretical description of symmetrical components of the experimental curves
\begin{equation} \label{symm}
\rho_{s}({\bm B}) = \frac{\rho({\bm B}) + \rho(-{\bm B})}{2}\:,\:J_{s}({\bm B}) = \frac{J({\bm B}) + J(-{\bm B})}{2}\:,
\end{equation}
depicted in Fig. \ref{fig3}, (b) and (d), using the data of Fig. \ref{fig1}.

It is seen that the theory qualitatively reproduces the evolution of the curves $\rho_{s}({\bm B})$ and $J_s({\bm B})$ with the angle $\alpha$ from 0 (longitudinal field) to $90^{\circ}$ (transverse field). In the longitudinal field, these curves have a minimum at $B = 0$. As the field deviates from the longitudinal direction the polarization and intensity sink down and, simultaneously, a narrow maximum rises above the flat minimum in the vicinity of the point $B = 0$, so that each curve is characterized by one maximum and two minima, one on the left and right. In the transverse field the minima disappear and there is only a maximum at $B = 0$, due to the Hanle effect. Although, on the whole, the calculated dependences in Fig. \ref{fig3} satisfactorily describe the vertical evolution of the experimental curves, there is a significant discrepancy between theory and experiment for widths of the minima and maxima. This discrepancy may be connected with the usage of a simplified model of the spin-dependent recombination which does not take into account a more complex kinetics for defects with the nuclear spin $I = 3/2$.
\begin{figure}
\includegraphics[width=7.0cm,angle=0]{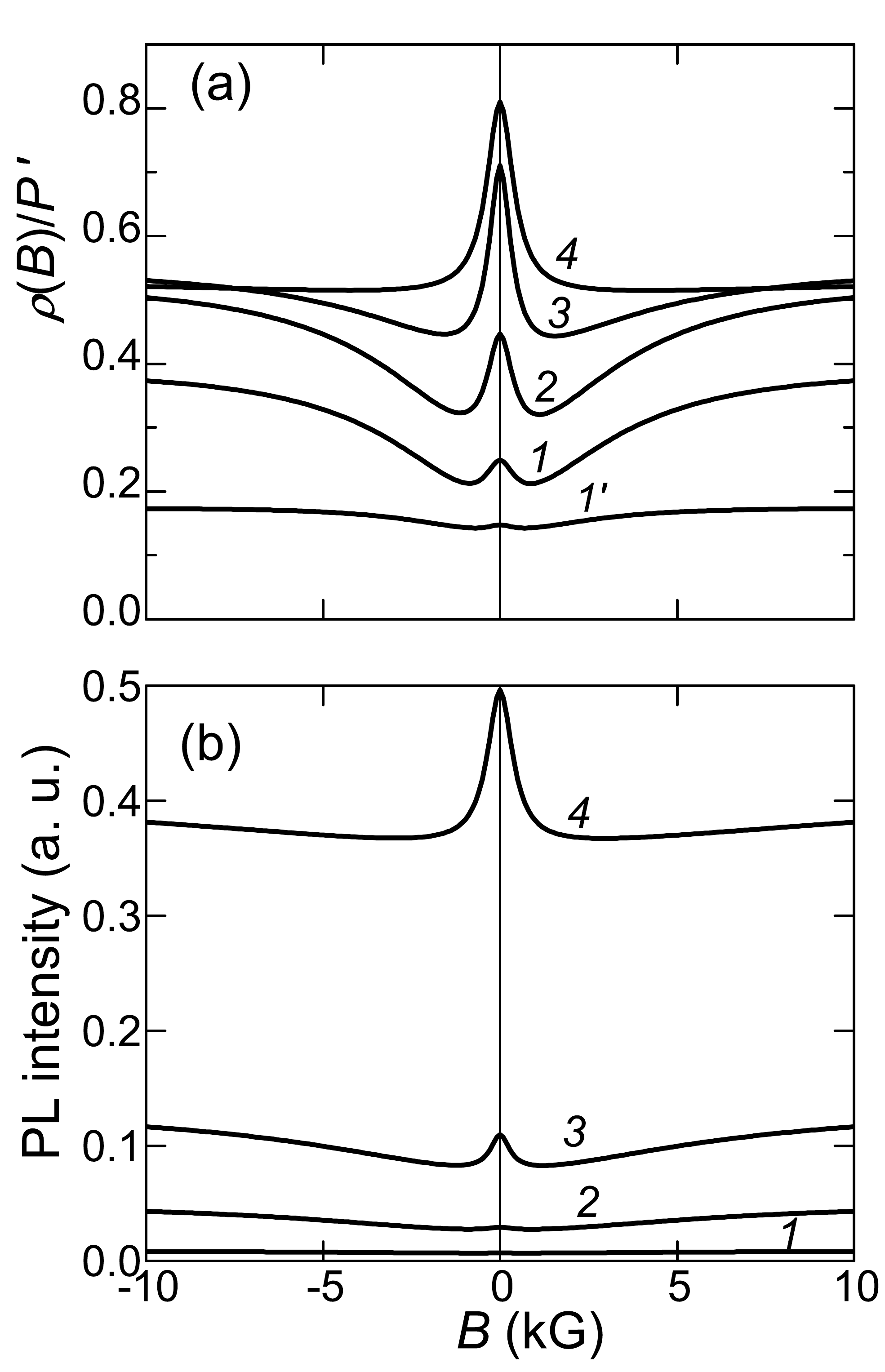}
\caption{\label{fig4} Magnetic-field dependences of the circular polarization degree (a) and intensity (b) of
photoluminescence calculated for the magnetic-field tilt angle $\alpha$ =
45$^{\circ}$ and different pump powers W:  1$^{\prime}$ -- 4
mW, 1 -- 20 mW, 2 -- 40 mW, 3 -- 60 mW, 4 -- 100 mW. All other model parameters
are the same as those in Fig.~\ref{fig2}.}
\end{figure}

The behavior of electron optical orientation in the oblique field with two minima and a sharp maximum in the middle of the curves $\rho({\bm B})$ and $J({\bm B})$ unambiguously evidences the strong coupling between the spin of a localized electron and the nuclear spin of the defect. We stress the distinction between the electron-nuclear system under consideration and the system (investigated in the 1970s and 1980s) of shallow-donor-bound electrons experiencing the contact hyperfine interaction with nuclei of the main lattice of the semiconductor \cite{OO,Safarov1974,FTT1981}:
in the latter case the interaction of the electron spin with the spin of a single nucleus is weak but, because of a large number of nuclei ($\sim$$10^5$) enveloped by the electron cloud of localized state, the electron-nuclear system turns to be strongly coupled. In this system an important role is played by cooling of the nuclear spin subsystem; as a consequence, in addition to the central peak at ${\bm B} = 0$, the curve of magnetic depolarization of the photoluminescence in the oblique field contains two additional peaks, one of which is adjacent to the central peak, and the second is shifted towards strong fields. Depending on the relative signs of electron and nuclear $g$ factors, the additional peaks are located on opposite sides or one side relative to the central peak; they arise at the values of the magnetic field at which the nuclear field compensates the external field.

Figure \ref{fig4} illustrates the variation of the magnetic-field dependences $\rho({\bm B})$ and $J({\bm B})$ calculated at $\alpha = 45^{\circ}$ with increasing the pump power. For a very low intensity of the circularly polarized excitation, the spins of electrons on paramagnetic centers remain unpolarized and the magnetic field has practically no influence on the PL polarization or intensity. Starting from $W$ = 20 mW for the curve $\rho({\bm B})$ and $W$ = 60 mW for the curve $J({\bm B})$, a narrow maximum at ${\bm B} = 0$ and two minima on each side are being formed. With a further increase in the power $W$, the height of the maximum increases, and at the pump of $W$ = 100 mW only a single maximum is left on the curve $\rho({\bm B})$. To explain this result, we recall that in the approximate description of the curve $\rho({\bm B})$ the side minima arise as a result of the superposition of two Lorentzian contours, normal and inverted. The inverted contour appears due to suppression of the electron-nuclear spin-spin interaction by the longitudinal component of the magnetic field. As mentioned in Ref. \cite{PRB2015}, at high intensity of the exciting light the lifetime $\tau_c = (c_n n)^{-1}$ of the defect with one electron becomes so short that the uncertainty $\hbar/\tau_c = \hbar c_n n$ exceeds the hyperfine interaction constant $A$, consequently 
this interaction reduces and is negligible even in the absence of magnetic field. Therefore, for large values of $W$ the longitudinal component of the magnetic field has no effect on the electron spin polarization, see, e.g., Fig. 1 in Ref. \cite{PRB2015}.
\begin{figure}
\includegraphics[width=7.0cm,angle=0]{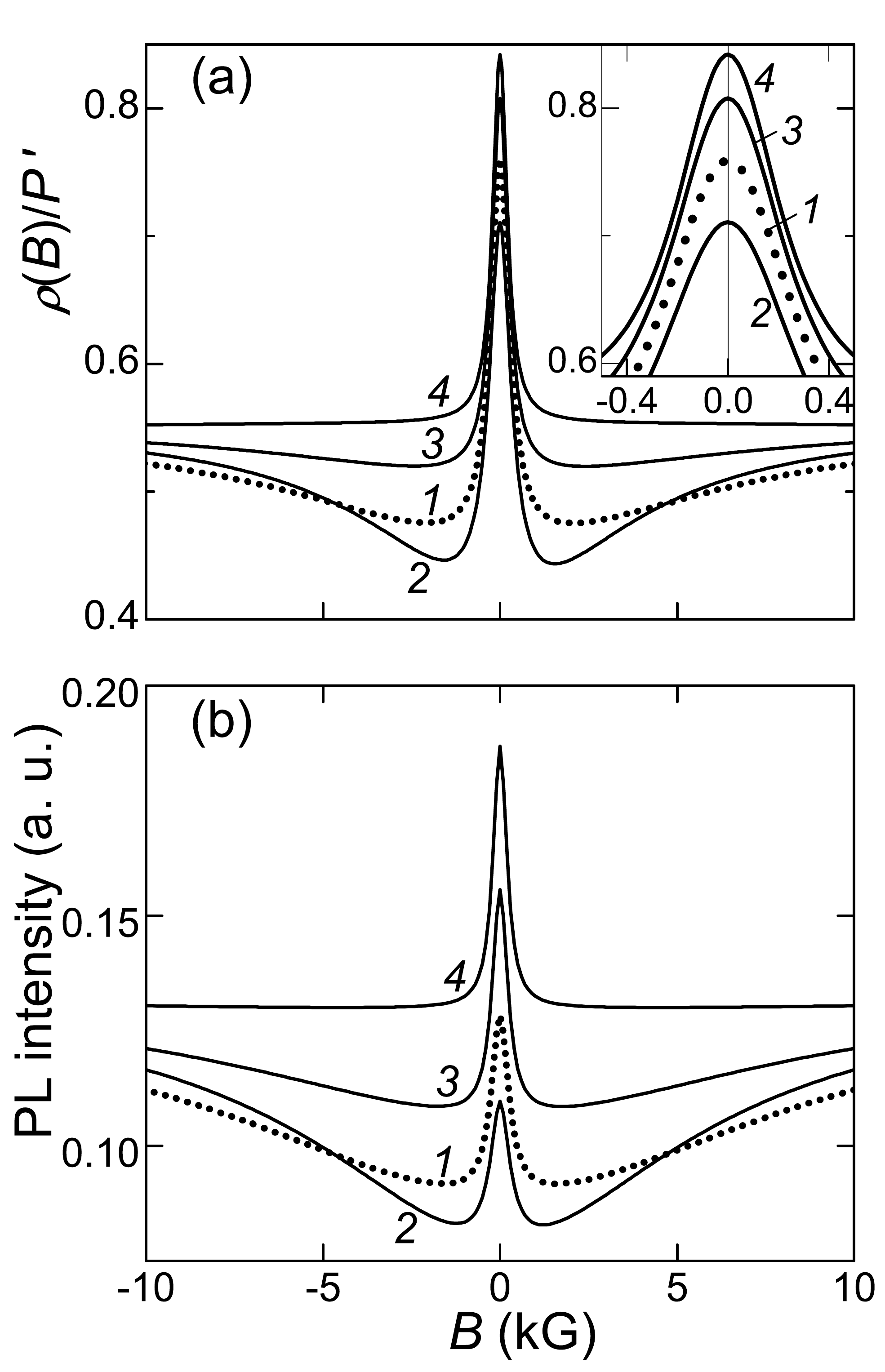}
\caption{\label{fig5} Magnetic-field dependences of the circular polarization degree (a) and intensity (b) of
photoluminescence calculated for the magnetic-field tilt angle $\alpha$ = 45$^{\circ}$, pump power $W$ = 60 mW
and four values of the coinciding nuclear spin-relaxation times $\tau_{n1} = \tau_{n2}$: 1 -- 15 ps, 2 -- 150 ps, 3 -- 1500 ps, 4 --
15000 ps. All other model parameters are the same as those in Fig. \ref{fig2}. Inset shows
parts of the same dependences $\rho({\bm B})$ in an enlarged scale.}
\end{figure}

The sensitivity of magnetic-field dependences $\rho({\bm B})$ and $J({\bm B})$ to the choice of the nuclear relaxation times is illustrated in Fig. \ref{fig5}. For this purpose we have chosen the particular values $\alpha$ = 45$^{\circ}$ and $W$ = 60 mW, set $\tau_{n1} = \tau_{n2} $ and performed calculations for four different values of these times. As shown in Ref. \cite{PRB2015}, for very long times of nuclear relaxation under the stationary optical excitation, the nuclear spins are polarized in such a way that the backward influence of the nuclei on the electron spin polarization vanishes, see, e.g., Fig. 3(a) in \cite{PRB2015}. In this case, the switching on of a longitudinal magnetic field does not affect $\rho$ and $J$. Therefore, in the curves 4 in Fig. \ref{fig5} calculated at $\tau_{n1}=\tau_{n2}$ = 150000 ps, there exists a pronounced maximum and there are no minima. The height of normal Lorenzian and the depth of inverted one are comparable within 15 ps $< \tau_{n1} = \tau_{n2} <$ 150 ps. In the strong magnetic field, the hyperfine interaction is broken off, the component of spin polarization of localized electrons ${\bm S}_{c \perp} \perp {\bm B}$ is suppressed, and all four curves in Fig. \ref{fig5}(a) or Fig. \ref{fig5}(b) converge to each other. We have derived a formula for the PL circular polarization after the magnetic-field-induced suppression of the transverse component of the polarization ${\bm S}_{c \perp}$. To this end, we have neglected the Hanle effect on the free electrons setting $g= 0$ and calculated $\rho$ in the absence of hyperfine interaction. Here is the final result
\begin{equation} \label{infty}
\rho(\alpha; {\bm S}_{c \perp}=0, A = 0) = \frac{P' P_i G}{n} \left( T' \cos^2{\alpha} + T \sin^2{\alpha}\right)\:,
\end{equation} 
where the times $T, T'$ are defined by 
\[
\frac{1}{T} = \frac{1}{\tau} + \frac{1}{\tau_s}\:\:,\:\:\frac{1}{T'} = \frac{1}{T} - \frac{T_c}{\tau \tau_c}\:\:,\:\:\frac{1}{T_c} = \frac{1}{\tau_c} + \frac{1}{\tau_{sc}}\:,
\]
$\tau = (c_n N_1)^{-1}$, $\tau_c = (c_n n)^{-1}$, and other parameters are introduced in Eqs. (\ref{Lall}) and (\ref{p'}). The approximate formula (\ref{infty}) for $\rho$ in the magnetic field, strong enough to suppress the electron-nuclear spin-spin interaction but weak for the manifestation of the Hanle effect on the photoelectrons in the conduction band, is in good agreement with the numerical solution of the set (\ref{Lall}).
\section{Conclusion}
We have carried out an experimental and theoretical study of optical orientation and spin-dependent Shockley-Read-Hall recombination in a semiconductor in an oblique magnetic field at normal incidence of the circularly polarized radiation on the sample surface. The experiments have been performed at room temperature in the GaAs$_{1-x}$N$_x$ alloys where the deep paramagnetic centers responsible for the spin-dependent recombination are Ga$^{2+}$ self-interstitial defects. We have successfully represented the experimental magnetic-field dependences of the photoluminescence circular polarization $\rho({\bm B})$ and intensity $J({\bm B})$ as a superposition of two Lorentzian contours, normal and inverted, with their half-widths at half-height (half-depth) differing remarkably and equal to $\sim$100 G and $\sim$1000 G, respectively. Such kind of dependence $\rho({\bm B})$ or $J({\bm B})$ is related with a change in the spin state of electrons localized on the paramagnetic centers. The normal (narrow) Lorentzian is caused by depolarization of the localized-electron spin polarization component perpendicular to the direction of the external magnetic field (Hanle effect), whereas the inverted (broad) Lorentzian is caused by the suppression of hyperfine interaction of localized electron with the single nucleus of the defect and the elongation of the spin relaxation time of localized electrons. The relation between the height of one Lorentzian and depth of the other is determined by the field tilt angle $\alpha$. In the longitudinal field ($\alpha=0$) the normal Lorentzian is absent, and the inverted Lorentzian has the deepest minimum. The deviation of the magnetic field from the direction of excitation is accompanied by an appearance of a normal Lorentzian in the form of a narrow maximum at $B= 0$ superimposed on the background of the broad inverted Lorentzian. With the increasing inclination of the field the contribution of narrow Lorentzian increases and that of the broad one decreases, and at $\alpha = 90^{\circ}$ only the narrow Lorentzian remains. In contrast to the hyperfine interaction of an electron bound to a shallow donor with a large number of nuclei of the crystal lattice, in the studied electron-nuclear system the variation of the magnetic-field tilt angle is not followed by an appearance, in the magnetic-field dependence of the electron polarization, of an additional narrow peak shifted with respect to the point $B = 0$. This result demonstrates that, in the GaAsN alloy, the hyperfine interaction of a localized electron with the single nucleus of the paramagnetic center keeps being strong even at room temperature. For a theoretical description of the experiment, we have used a model of spin-dependent recombination through deep paramagnetic centers with the nuclear momentum $I = 1/2$, previously developed by us for the longitudinal field, and generalized it to an arbitrary angle of the magnetic field orientation. Since in case of the oblique field all components of the spin-density matrix for the defects with one or two electrons are different from zero, in the modified model, as compared with the theory for the longitudinal field, the number of equations increases from 11 to 25. The calculated theoretical dependences $\rho({\bm B})$, $J({\bm B})$ agree with the approximate description of the experimental curves in the form of two Lorentzians, a second narrow shifted contour does not appear as well.

\acknowledgments{This research was partially supported by the Government of Russian Federation (project
14.Z50.31.0021) and the grants 14-02-00959 and 15-52-12012 of the Russian Foundation for Basic Research. We are grateful to T. Amand and A.Yu. Shiryaev for fruitful discussions.}

\end{document}